\documentclass[preprint,aps,eqsecnum]{revtex4}
\usepackage{epsfig}
\newcommand{\be}{\begin{eqnarray}}
\newcommand{\ee}{\end{eqnarray}}

\begin{document}
%\preprint{DO-TH ----}
%\preprint{hep-ph/0306275}
\vspace*{1cm}
\title
{Impact Parameter Dependent Parton Distributions for a Relativistic
Composite System}
\author{\bf D. Chakrabarti}   
\email{dipankar@phys.ufl.edu}
\affiliation{Department of Physics, University of Florida, Gainesville,
FL-32611-8440, USA}
\author{\bf A. Mukherjee}\email{asmita@lorentz.leidenuniv.nl}
\affiliation{ Institute Lorentz, University of Leiden, 2300 RA Leiden, 
Netherlands}

\date{\today\\[2cm]}

\begin{abstract}
We investigate the impact parameter dependent parton distributions for a
relativistic composite system in light-front framework. We take an effective
two-body spin-$1/2$ state, namely  an electron dressed with a photon in QED. We express
the impact parameter dependent parton distributions in terms of overlaps of
light-cone wave functions. We obtain the scale dependence of both fermion
and gauge boson distributions and show the distortion of the pdfs in the 
transverse space for transverse polarization of the state at one loop level.   

\end{abstract}
\maketitle

%%%%%%%%%%%%%%%%%%%%%%%%%%%%%%%%%%%%%%%%%%%%%%%%%%%%%%%%%%%%%%%%%%%%%%%%%%%
\section{Introduction}
%%%%%%%%%%%%%%%%%%%%%%%%%%%%%%%%%%%%%%%%%%%%%%%%%%%%%%%%%%%%%%%%%%%%%%%%%%%%%
Impact parameter dependent parton distributions $q(x,b^\perp)$ 
\cite{bur1,bur2} have been
introduced recently as a physical interpretation of generalized parton
distributions (GPDs) \cite{gpd} in terms of probability densities in the
impact parameter space. It is known that GPDs are off-forward matrix
elements of light-cone bilocal operators and they do no have a probabilistic
interpretation like ordinary parton distributions. $q(x,b^\perp)$ can be 
expressed as Fourier transform of GPDs
with respect to the transverse momentum transfer (when the longitudinal
momentum transfer is zero) and they give the
distribution of partons in the transverse position space. In fact they obey 
certain
positivity constraints and thus it is legitimate to associate the physical
interpretation as a probability density. This interpretation is not limited by
relativistic effects in the infinite momentum frame. Another interesting
aspect is that when the state is polarized in the transverse direction,   
the unpolarized impact parameter dependent pdf is distorted in the
transverse plane \cite{bur1}. This distortion can be linked to the
transverse single spin asymmetries observed in transversely polarized Lambda
production in $pp$ or $p {\bar p}$ collisions. Recently it has been shown in
the framework of the scalar diquark model that the impact parameter space
asymmetry together with the final state interaction of the active quark gives 
rise to the Sivers effect \cite{siv} in momentum space \cite{bur4}.

Impact parameter dependent pdfs have been investigated in chiral
quark model for the pion \cite{model1}, in the constituent quark model 
\cite{model2}, 
in terms of a power-law ansatz of the light-cone wave function 
for the pion \cite{model3}, in the context of investigating the color transparency
phenomena \cite{model4}, in the transverse lattice framework for the pion  
\cite{model5} and on the lattice \cite{lattice}. In this work, we
investigate both fermion and gauge boson distributions for a relativistic composite system in the light-front
framework, taking into account the correlation between different Fock
components of the light-cone (or light-front) wave function in light-front
gauge. It is well known that the 
light-front formalism is especially 
suitable to investigate many aspects of relativistic bound states because of
the Galilean transverse boost operators and the triviality of the vacuum
\cite{wilson} when a cutoff is imposed on the longitudinal momentum. We take an effective
spin ${1\over 2}$ system, namely an electron \cite{impact}, dressed with a photon in QED.
It is known that such a model is self consistent and has been used
to investigate the helicity structure of a composite relativistic system
\cite{brod1}. The state can be expanded in Fock space in terms of light-cone
wave functions. The light-cone wave functions in this case can be obtained
from perturbation theory, and thus their correlations are known at a certain
order in the coupling constant. However, their general form provides a template for
parametrizing the more realistic composite system, namely the hadronic wave
function \cite{brod1}. Instead of an electron, one can also consider a state like a
dressed quark in perturbative QCD. In fact this
can be thought of as a field theoretic parton model where the partons, or
the quarks and gluons, have mass, intrinsic transverse momenta and they
interact. Previous studies have shown that this gives an intuitive picture 
of the DIS structure functions 
and scaling violations \cite{hari} and is suitable to address issues
related to the 
spin and orbital angular momentum of the nucleon \cite{oam,hari1}. 
Such a state has also been used to investigate the twist three
GPDs in terms of overlaps of light-cone wave functions \cite{mvh}. 
The aim of the
present work is to extend these studies to the impact parameter dependent
fermion and gauge boson pdfs in order to obtain a qualitative behavior in
the impact parameter space within the framework of perturbation theory.        

The plan of the paper is as follows. In section II, we give the definition of
the impact parameter dependent pdf, In sections III and IV we calculate the
fermion and gauge boson distributions respectively. Summary and conclusions
are given in section V. 
%%%%%%%%%%%%%%%%%%%%%%%%%%%%%%%%%%%%%%%%%%%%%%%%%%%%%%%%%%%%%%%%%%%%%%%%%%%%%%%
\section{Impact Parameter Dependent Parton Distributions}
%%%%%%%%%%%%%%%%%%%%%%%%%%%%%%%%%%%%%%%%%%%%%%%%%%%%%%%%%%%%%%%%%%%%%%%%%%%%%%%
Impact parameter dependent pdfs are defined by considering a transversely
localized state \cite{bur1}:
\be
\mid P^+, R^\perp=0, \lambda \rangle= N\int {d^2P^\perp\over (2 \pi)^2}
\mid P^+, P^\perp, \lambda\rangle,
\ee
where $ \mid P^+, P^\perp, \lambda\rangle$ is the light cone helicity
eigenstate with momentum $P$ and helicity $\lambda$, $N$ is a normalization
factor. For a  state with total momentum $P^+$ the transverse center of
momentum $R^\perp$ is defined as,
\be
R^\perp= {1\over 2 P^+} \int dx^- d^2x^\perp \Theta^{++} x^\perp,
\ee
where $\Theta^{\mu \nu}$ is the energy-momentum tensor.  

The states $\mid P^+, P^\perp, \lambda\rangle$ can be thought of as helicity
eigenstates in the infinite momentum frame \cite{bur1}. Due to the fact that
the light-front transverse boost $B^\perp$ behaves similar to the
non-relativistic Galilean Boost generators, it can be shown that the states
$\mid P^+, R^\perp=0, \lambda \rangle$ defined above are simultaneous
eigenstates of the longitudinal light cone momentum $P^+$, transverse
position operator $R^\perp$ and light cone helicity $J^z$. The transverse
position operator $R^\perp$ in fact is related to $B^\perp$ : $
R^\perp=-{1\over P^+} B^\perp$.
 
The impact parameter dependent pdf is defined as,
\be
q(x,b^\perp)=  \langle P^+, R^\perp=0^\perp, \lambda
\mid O_q(x,b^\perp) \mid P^+, R^\perp=0^\perp,\lambda \rangle
\label{impact}
\ee
with
\be
O_q(x, b^\perp)= \int {dy^- \over 4 \pi} {\bar \psi} (-{y^-\over 2}, b^\perp)
\gamma^+ \psi({y^-\over 2}, b^\perp) e^{{i\over 2} x P^+ y^-},
\ee
$b^\perp$ is the impact parameter, which is the transverse distance
of the active quark from the center of mass. We have taken the light-front
gauge, $A^+=0$. Instead of the fermion
operator, one can also have a gauge boson operator 
\be
O_g(x, b^\perp)= \int {dy^- \over 4 \pi} F^{+ \nu}(-{y^-\over 2}, b^\perp)
{F^+}_\nu({y^-\over 2}, b^\perp) e^{{i\over 2} x P^+ y^-}
\ee
to define the gauge boson pdf $g(x,b^\perp)$ similar to Eq. (\ref{impact}).

It can be shown that $ q(x, b^\perp)$ can be expressed as a Fourier
transform of the GPD $H_q(x, 0, \Delta^2)$ \cite{bur1} :
\be
q(x,b^\perp)= {\mathcal H}_q (x,b^\perp) = \int {d^2 \Delta^\perp\over (2 \pi)^2}  
e^{-i b^\perp. \Delta^\perp} H_q(x, 0, \Delta^2).
\ee
where $ \Delta^2$ is the total momentum transfer and it is purely
transverse. One can obtain a similar relation for the gauge boson
counterpart.
%%%%%%%%%%%%%%%%%%%%%%%%%%%%%%%%%%%%%%%%%%%%%%%%%%%%%%%%%%%%%%%%%%%%%%%%%%%%%%
\section{Fermion  Distributions}
%%%%%%%%%%%%%%%%%%%%%%%%%%%%%%%%%%%%%%%%%%%%%%%%%%%%%%%%%%%%%%%%%%%%%%%%%%%%%%
We first calculate the off-forward matrix elements, parametrized in terms of
the twist-two GPSs $ H_q({x},\Delta^2)$ and $ E_q({x},\Delta^2) $ in the 
standard way \cite{ji} and using the light cone
spinors \cite{ped}. We have, 
\be
\int {dz^-\over {8 \pi}} e^{{i\over 2}{x}{\bar
P^+} {z^-}}\langle P' \uparrow  \mid \bar \psi (-{ z^-\over 2}) 
\gamma^+\psi ({ z^-\over 2})
\mid P \uparrow \rangle = H_q({x},\Delta^2), 
\label{def}
\ee
where $H(x, \Delta^2)= H(x,0, \Delta^2)$.
The momentum of the initial state
is $P^\mu$ and that of the final state is $P'^\mu$. The average momentum
between initial and final state is then ${\bar P}^\mu={P^\mu+P'^\mu\over
2}$. 
The momentum transfer is given by $\Delta^\mu=P'^\mu-P^\mu$, $ P'_\perp =
-P_\perp
={\Delta_\perp\over 2}$. We take the skewedness $\xi=0$, in other words, the
momentum transfer is purely transverse. We exclude the impact parameter space
representation of GPDs for non-zero skewedness
\cite{diehl} from our analysis here. 

We take the state 
$ \mid P, \sigma \rangle$ to be a dressed electron
consisting of bare states of an electron and an electron plus
a photon :
%%%%%%%%%%%%%%%%%%%%%%%%%%%%%%%
%\begin{eqnarray}
%\mid P, \sigma \rangle && = \phi_1 b^\dagger(P,\sigma) \mid 0 \rangle
%\nonumber \\  
%&& + \sum_{\sigma_1,\lambda_2} \int 
%{dk_1^+ d^2k_1^\perp \over \sqrt{2 (2 \pi)^3 k_1^+}}
%\int 
%{dk_2^+ d^2k_2^\perp \over \sqrt{2 (2 \pi)^3 k_2^+}}   
%\sqrt{2 (2 \pi)^3 P^+} \delta^3(P-k_1-k_2) \nonumber \\
%&& ~~~~~\phi_2(P,\sigma \mid k_1, \sigma_1; k_2 , \lambda_2) b^\dagger(k_1,
%\sigma_1) a^\dagger(k_2, \lambda_2) \mid 0 \rangle.
%\label{eq2}    
%\end{eqnarray} 
%%%%%%%%%%%%%%%%%%%%%%%%%%%%%%%%%%%%
\begin{eqnarray}
\mid P, \sigma \rangle && = {\cal N}\Big [ b^\dagger(P,\sigma) \mid 0 \rangle
\nonumber \\  
&& + \sum_{\sigma_1,\lambda_2} \int 
{dk_1^+ d^2k_1^\perp \over \sqrt{2 (2 \pi)^3 k_1^+}}
\int 
{dk_2^+ d^2k_2^\perp \over \sqrt{2 (2 \pi)^3 k_2^+}}   
\sqrt{2 (2 \pi)^3 P^+} \delta^3(P-k_1-k_2) \nonumber \\
&& ~~~~~\phi_2(P,\sigma \mid k_1, \sigma_1; k_2 , \lambda_2) b^\dagger(k_1,
\sigma_1) a^\dagger(k_2, \lambda_2) \mid 0 \rangle \Big ].
\label{eq2}    
\end{eqnarray} 

Here $a^\dagger$ and $b^\dagger$ are bare photon and electron
creation operators respectively and $\phi_2$ is the
multiparton wave function. It is the probability amplitude to find 
one electron plus photon inside the dressed electron state.

We introduce Jacobi momenta $x_i$, ${q_i}^\perp$ such that $\sum_i x_i=1$ and
$\sum_i {q_i}^\perp=0$.  They are defined as
\be
x_i={k_i^+\over P^+}, ~~~~~~q_i^\perp=k_i^\perp-x_i P^\perp.
\ee
%%%%%%%%%%%%%%%%%%%%%%%%%%%%%%%%%
%Also, we introduce the wave functions,  
%\be
%\psi_1=\phi_1, ~~~~~~~~~~~\psi_2(x_i,q_i^\perp)= {\sqrt {P^+}} \phi_2
%(k_i^+,{k_i}^\perp);
%\ee
%which are independent of the total transverse momentum $P^\perp$ of the
%state and are boost invariant.
%%%%%%%%%%%%%%%%%%%%%%%%%%%%%%%%%%%%%%%
Also, we introduce the wave function,  
\be
\psi_2(x_i,q_i^\perp)= {\sqrt {P^+}} \phi_2 (k_i^+,{k_i}^\perp);
\ee
which is independent of the total transverse momentum $P^\perp$ of the
state and  boost invariant.
The state is normalized as,
\be
\langle P',\lambda'\mid P,\lambda \rangle = 2(2\pi)^3
P^+\delta_{\lambda,\lambda'} \delta(P^+-{P'}^+)\delta^2(P^\perp-P'^\perp).
\label{norm}
\ee
The two particle wave function depends on the helicities of the electron and
photon. Using the eigenvalue equation for the light-cone Hamiltonian, this
can be written as \cite{hari},
\be
\psi^\sigma_{2\sigma_1,\lambda}(x,q^\perp)&=& -{x(1-x)\over
(q^\perp)^2+m^2 (1-x)^2}
{1\over {\sqrt {(1-x)}}} {e\over
{\sqrt {2(2\pi)^3}}} \chi^\dagger_{\sigma_1}\Big [ 2 {q^\perp\over
{1-x}}+{{\tilde \sigma^\perp}\cdot q^\perp\over x} {\tilde \sigma^\perp}
\nonumber\\&&~~~~~~~~~~~~~~~~~~
-i m{\tilde \sigma}^\perp {(1-x)\over x}\Big ]\chi_\sigma
\epsilon^{\perp *}_\lambda .
\label{psi2}
\ee
$m$ is the bare mass of the electron, $\tilde \sigma^2 = -\sigma^1$ and
$\tilde \sigma^1= \sigma^2$. In our case,
$\Delta^2=-(\Delta^\perp)^2$ as $\xi=0$. 
%$\psi_1$ actually gives the normalization of the state \cite{hari}:
${\cal N}$  gives the normalization of the state \cite{hari}:
\be
{\cal N}^2=1-{\alpha\over {2 \pi}} 
\int_\epsilon^{1-\epsilon} dx \Big [{{1+x^2}\over {1-x}}log
{Q^2\over \mu^2}-{1+x^2\over 1-x}+(1-x) \Big ],
\label{c5nq}
\ee
within order $\alpha$.  Here $\epsilon$ is a small cutoff on $x$. We have
taken the cutoff of the transverse momenta to be $Q^2$, $\mu^2$ is a scale
which we have taken to be $ m^2 << \mu^2 << \Lambda^2$, it mimics the
factorization scale in QCD separating hard and soft dynamics \cite{hari}. 
The above expression is derived using Eqs (\ref{norm}), (\ref{eq2}) and
(\ref{psi2}).

Contribution  to $H_q(x, \Delta^2)$ comes from one particle and two-particle sectors.
The one-particle sector contributes only when
${x}=1$ and is a delta function. This receives correction upto order $\alpha$
from the normalization condition of the state.
The contribution from the two-particle sector can be written as an overlap
of wave functions \cite{overlap,mvh},
\be
{\cal N}^2 \sum \int d^2q^\perp \psi^*_2(x,
q^\perp+(1-x) \Delta^\perp) \psi_2(x, q^\perp).
\label{matrix+}
\ee

Using the expression Eq. (\ref{psi2}) we get 
\be
H_q(x, \Delta^2)&=&{\cal N}^2 \delta (1-x) \nonumber \\&&
\!\!\!\!\!\!\!\!\!\!+ {e^2\over (2 \pi)^3} {\cal N}^2 \Big [ 
\int d^2 q^\perp {{{1+x^2\over 1-x} (q^\perp)^2 +m^2(1-x)^3}\over
{((q^\perp)^2+m^2 (1-x)^2) ((q^\perp+(1-x) \Delta^\perp)^2 +m^2
(1-x)^2)}}\nonumber\\&&\!\!\!\!\!\!\!\!\!\!\!+
(1+x^2)\int d^2q^\perp {q^\perp .\Delta^\perp\over 
{((q^\perp)^2+m^2 (1-x)^2) ((q^\perp+(1-x) \Delta^\perp)^2 +m^2 (1-x)^2)}} \Big
]. 
\ee     
One can write this as,
\be
H_q(x, \Delta^2)&=& {\cal N}^2 \Bigg \{ \delta (1-x) 
+ {e^2\over (2 \pi)^3} \Bigg [ \Big ({1+x^2\over 1-x} \Big ) \Big [ \int {d^2q^\perp\over L_2} 
\nonumber\\&&-m^2 (1-x)^2 \int {d^2 q^\perp\over L_1L_2}\Big ] +m^2 (1-x)^3 \int {
d^2 q^\perp\over L_1L_2} +{1+x^2\over 2 (1-x)} \Big [ 
\int {d^2q^\perp\over L_1}\nonumber\\&&-\int {d^2q^\perp\over L_2}-(1-x)^2
(\Delta^\perp)^2 \int {d^2q^\perp\over L_1 L_2} \Big ] \Bigg ] \Bigg \},
\label{hqtot}
\ee   
where $L_1=(q^\perp)^2+m^2(1-x)^2$ and $L_2=(q^\perp+(1-x)
\Delta^\perp)^2+m^2 (1-x)^2$. 
In the forward limit using the normalization 
condition of the state we get,
\be
H_q(x,0)&=&\delta (1-x)+ {\alpha\over 2\pi} \Bigg \{ \Big [ {1+x^2\over 1-x}log
{Q^2\over \mu^2} -{1+x^2\over 1-x} +1-x \Big ] \nonumber\\&&- \delta (1-x) 
\Big [\int dy
\Big ( {1+y^2\over 1-y} log {Q^2\over \mu^2}-{1+y^2\over 1-y}+1-y \Big )
\Big ] \Bigg \}.
\label{for}
\ee  
Integrating over $x$, one gets $\int_0^1 dx H_q(x,0)= F_1(0)=1$, where
$F_1(0)$ is the form factor at zero momentum transfer. 
Replacing $\alpha$ by $\alpha_s C_f$ and neglecting the mass, eq.
(\ref{for}) reduces to  the quark distribution
function of a dressed quark and the coefficient of the logarithmic term
gives the splitting function $P_{qq}$ \cite{hari}.  At
non-zero $\Delta^\perp$, in the limit $x \rightarrow 1$, we have
\be
H_q(x, \Delta^2) \rightarrow \delta (1-x)+ {\alpha\over 2 \pi} log
{Q^2\over \mu^2} \Big [ {1+x^2\over (1-x)_+}+{3\over 2} \delta (1-x)\Big ].
\ee
The plus '+' prescription is defined in the usual way. There is no singularity 
at $x \rightarrow 1$ and $H_q(x, \Delta^2)$ is
independent of $\Delta^\perp$ at $x \rightarrow 1$.
The impact parameter
dependent parton distribution is obtained by taking a Fourier transform with
respect to $\Delta^\perp$. At $x \rightarrow 1$ , we
get
\be
{\mathcal H}_q(x,b^\perp)&=& \int {d^2 \Delta^\perp\over (2 \pi)^2}  
e^{-i b^\perp. \Delta^\perp} H_q(x,
\Delta^2)\nonumber\\&&
=\delta^2 (b^\perp)\Big \{  \delta (1-x)+ {\alpha\over 2 \pi} log
{Q^2\over \mu^2} \Big [ {1+x^2\over (1-x)_+}+{3\over 2} \delta (1-x)\Big ]
\Big \}.\label{hq}
\ee 
as expected because in this limit the electron carries all the momentum 
and the transverse width of
the impact parameter dependent pdf vanishes \cite{bur1}. 

For $ x $ not equal to $1$, $ {\mathcal H}_q(x,b^\perp) $ has nontrivial 
 $b^\perp$ dependency which comes from the (finite)
mass terms as well as the $\Delta^\perp$ terms. The $q_\perp$ integration from which 
we get the scale dependency in the above expression (Eqn. \ref{hq}) cannot be done
analytically for this case. Here
we do not plot the scale dependent $ {\mathcal H}_q(x,b^\perp) $. 

Next we consider the helicity flip part of the matrix element :
\be
\int {dy^-\over 8 \pi} &&e^{{i\over 2} P^+ y^-x}\langle P+\Delta, \uparrow
\mid {\overline \psi}({-y^-\over 2}) \gamma^+ \psi ({y^-\over 2})\mid P, 
\downarrow \rangle =
{e^2\over (2 \pi)^3} x (1-x)^2 (-i m) (-i \Delta^1-\Delta^2)\nonumber\\&&
    \int
{d^2 q^\perp \over {((q^\perp)^2+m^2 (1-x)^2) ((q^\perp+(1-x) \Delta^\perp)^2 
+m^2 (1-x)^2)}} \nonumber\\&&=-{E_q\over 2 m} (\Delta^1-i \Delta^2).
\ee
$m$ is the renormalized mass of the electron. In the limit $\Delta^\perp =0$ 
we get
\be
E_q(x,0)= {\alpha\over\pi} x.
\ee
Integration over $x$ gives the Schwinger value for the anomalous magnetic
moment of an electron in QED : 
\be
\int_0^1 E_q dx = F_2(0) = {\alpha\over 2 \pi}.
\ee
This result can also be obtained by directly calculating the matrix element
of the current operator in the light-front framework \cite{brod1}.

\begin{center}  
%\hspace{-0.3cm}
\parbox{8cm}{\epsfig{figure=Fig1a_v2.eps,width=7.7 cm,height=7 cm}}\
\
%\hspace{0.3cm}
\parbox{8cm}{\epsfig{figure=Fig1b_v2.eps,width=7.5 cm,height=7 cm}}\
\
\end{center}   
\vspace{0.2cm} 
\begin{center} 
\parbox{14.0cm}
{{\footnotesize
Fig. 1: (Color online) (a) ${\mathcal E}_q (x, b^\perp)$ vs $b^\perp$ for three different values
of $x$; (b) Derivative of ${\mathcal E}_q (x, b^\perp)$ with respect to
$b^x$ as a function of $b^\perp$. }}
\end{center}

For non-zero $\Delta^\perp$, $E_q$ becomes,
\be
E_q(x, \Delta^2) = {\alpha\over \pi} m^2 \int_0^1 d \beta {x \over
( \beta (1-\beta) (\Delta^\perp)^2 +m^2)},    
\label{eq}
\ee
$\beta $ is the Feynman parameter. The scale dependence in this case is
suppressed and we can integrate over the full $\mid q^\perp \mid $ range, 
from $0$ to $\infty$.

Taking the Fourier transform, we get
\be
{\mathcal E}_q(x, b^\perp)&=&\int {d^2 \Delta^\perp\over (2 \pi)^2}  
e^{-i b^\perp. \Delta^\perp} E_q(x,-(\Delta^\perp)^2)\nonumber\\&&
= \int {\Delta d \Delta\over (2 \pi)^2} d \theta E_q(x, -\Delta^2) 
e^{-i b \Delta cos \theta}
\nonumber\\&&=  {1\over 2 \pi}\int_0^\infty \Delta d \Delta 
E_q (x, -\Delta^2) J_0(b \Delta).
\label{fteq}
\ee
For numerical calculation we use the real integral form of the Bessel
function :
\be
J_0(b \Delta)= {1\over \pi} \int_0^\pi cos (b \Delta sin \theta) d \theta.
\ee

If we take a state polarized in the ${\hat y}$ direction (in the infinite
momentum frame) and whose center of momentum is at the origin,
\be
\mid P^+, R^\perp=0,{\hat y} \rangle = {1\over \sqrt 2} ( 
\mid P^+, R^\perp=0, \uparrow \rangle+i 
\mid P^+, R^\perp=0, \downarrow \rangle);
\ee
the unpolarized fermion distribution in the impact parameter space gets
distorted,
\be
q_{\hat y}(x,b^\perp) & =&  \langle P^+, R^\perp=0^\perp, {\hat y}
\mid O_q(x,b^\perp) \mid P^+, R^\perp=0^\perp, {\hat y} \rangle \nonumber\\
&&= {\mathcal H }_q(x,b^\perp) +{1\over 2 m}
{\partial \over \partial b^x} {\mathcal E}_q (x, b^\perp). 
\ee
The distortion is directly related to ${\mathcal E}_q (x,b_\perp)$. Fig. 1
(a) shows the helicity flip contribution ${\mathcal E}_q (x, b^\perp)$ as a
function of $b^\perp$ for three different values of $x$. We have plotted for
positive $b^\perp$. ${\mathcal E}_q (x, b^\perp)$ is a smooth function of
$b^\perp$ in the range shown and it increases as $b^\perp$ decreases. Also,
it increases linearly with $x$, as is clear from Eq.(\ref{eq}). We have taken
the overall normalization ${\alpha\over 2 \pi}=1$ in order to study the
qualitative behavior and $m=0.5$. ${\mathcal E}_q (x, b^\perp)$ has a maximum at
$b^\perp=0$. Fig. 1 (b) shows the derivative of ${\mathcal E}_q (x, b^\perp)$
with respect to $b^x$, which gives the distortion of the pdf of an electron
at one loop 
in impact parameter space due to transverse polarization.  The integrand 
in this case contains $J_1(b \Delta )$ and it is highly
oscillatory. However, the integral converges 
for the $b^\perp$ range shown
in the plot. As ${\mathcal E}_q (x, b^\perp)$  is a smooth
function with maximum at $b^\perp=0$ its derivative for positive $b^\perp$
is negative. One can see that the distortion of the distribution in impact
parameter space increases
as $b^\perp$ decreases and for a given $b^\perp$ the distortion is higher in
magnitude for larger values of $x$. The distortion shifts the distribution
actually toward negative values of $b^\perp$.   
   
%%%%%%%%%%%%%%%%%%%%%%%%%%%%%%%%%%%%%%%%%%%%%%%%%%%%%%%%%%%%%%%%%%%%%%%%%%%%%%%%
\section{Gauge Boson Distributions}
%%%%%%%%%%%%%%%%%%%%%%%%%%%%%%%%%%%%%%%%%%%%%%%%%%%%%%%%%%%%%%%%%%%%%%%%%%%%%%%

From the definition,
\be
H_g(x, \Delta^2)= {1\over 8 \pi x P^+} \int dy^- e^{{i\over 2} P^+
y^- x} \langle P' , \uparrow \mid F^{+ \nu} (-{y^-\over 2}) F^+_\nu ({y^-\over
2}) \mid P, \uparrow \rangle;
\ee
we take the state to be a dressed electron as in Eq. (\ref{eq2}). 
Contribution in this case comes only from the two-particle sector, and we
get
\be
H_g(x, \Delta^2) &=& {\cal N}^2 \int d^2 q^\perp \psi_2^* (1-x,q^\perp)
\psi_2(1-x, q^\perp+(1-x) \Delta^\perp) \nonumber\\
&=&{\alpha\over (2\pi)^2} \Bigg \{ {(1+(1-x)^2)\over x} \int {d^2q^\perp
\over L_2} +m^2
x^3 \int {d^2 q^\perp\over L_1 L_2} +{1+(1-x)^2\over 2 x} \Big [ \int
{d^2 q^\perp\over L_1} \nonumber\\&&-\int {d^2q^\perp\over L_2}-(1-x)^2 
(\Delta^\perp)^2 \int {d^2q^\perp\over L_1 L_2} \Big ] \Bigg \},    
\ee
where $L_1=(q^\perp)^2 +m^2 x^2$ and $L_2=(q^\perp +(1-x) \Delta^\perp)^2+m^2
x^2$.
The scale dependency as before, comes from the limits of the $q^\perp$
integrations. Fourier transform of the above expression gives the impact
parameter dependent gauge boson pdf. Nontrivial $b^\perp$ dependence come
from the mass term and the $\Delta^\perp$ term. The other terms give
logarithmic dependence on the scale.
In the forward limit,
\be
H_g(x, 0)
= {\alpha\over 2 \pi} \Big [{ 1+(1-x)^2\over x}\Big ] log{Q^2\over \mu^2}. 
\ee

\begin{figure}  
%\hspace{-0.3cm}
\parbox{8cm}{\epsfig{figure=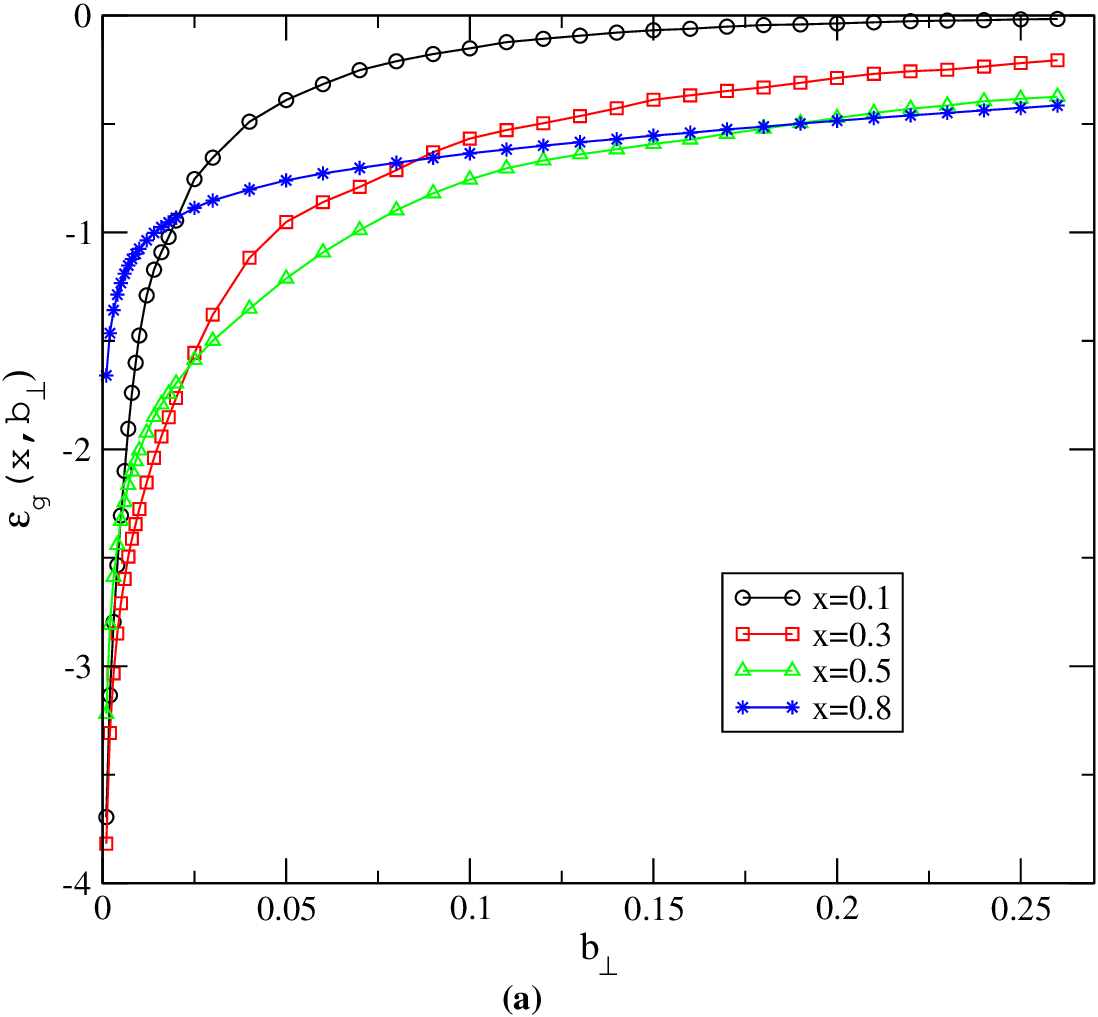,width=7.7 cm,height=7 cm}}\
\
%\hspace{0.3cm}
\parbox{8cm}{\epsfig{figure=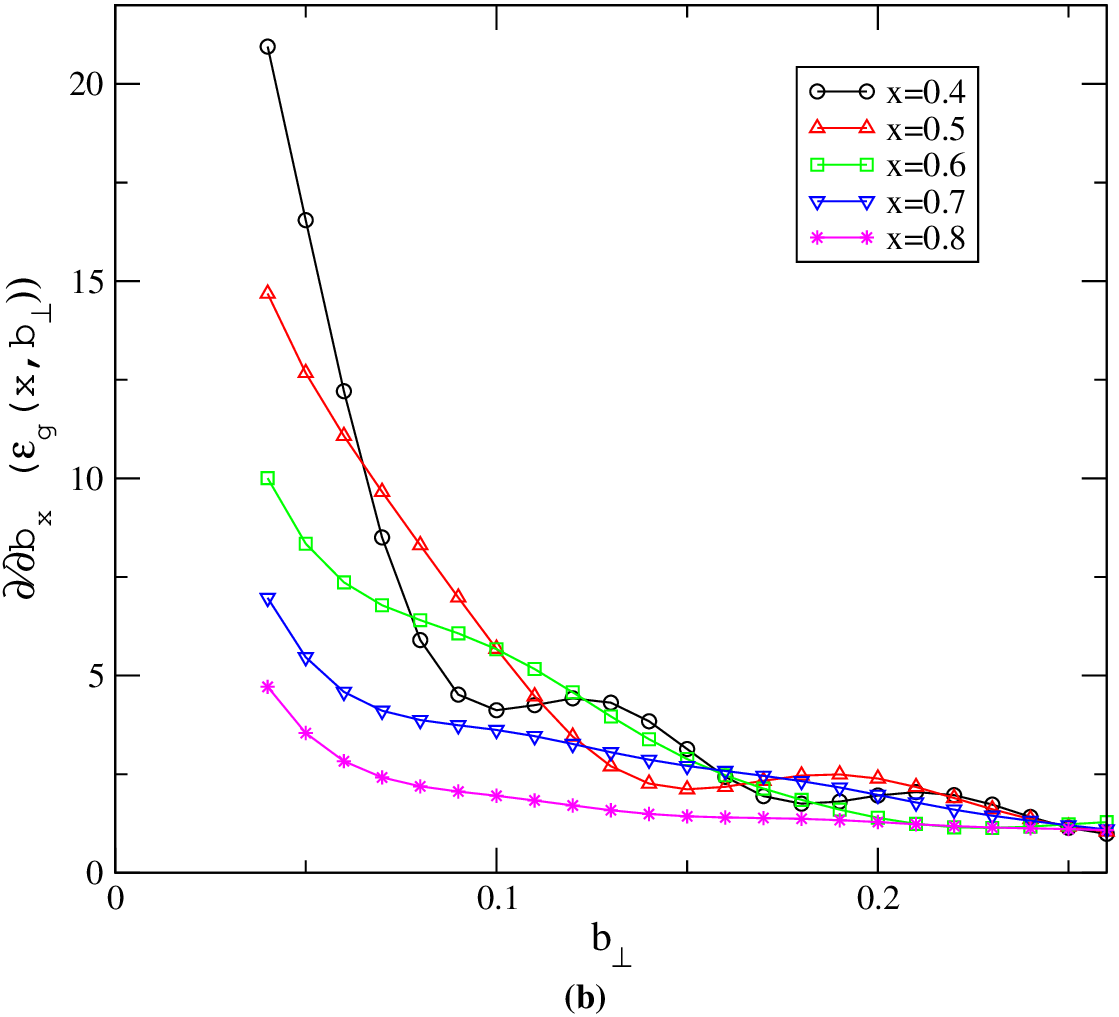,width=7.5 cm,height=7 cm}}\
\
\vspace{0.2cm} 
\begin{center} 
\parbox{14.0cm}
{{\footnotesize
Fig. 2: (Color online) (a) ${\mathcal E}_g (x, b^\perp)$ vs $b^\perp$ for four different values
of $x$; (b) Derivative of ${\mathcal E}_g (x, b^\perp)$ with respect to
$b^x$ as a function of $b^\perp$. }}
\end{center}
\end{figure}

Next we look at the helicity flip part of the matrix element. This can be
written as,

\be
\int {dy^-\over 8 \pi} e^{{i\over 2} P^+ y^-x}\langle P+\Delta, \uparrow
\mid F^{+ \nu}({-y^-\over 2}) F^+_\nu ({y^-\over                 
2}) \mid P, \downarrow \rangle
=-{E_g\over 2 m} (\Delta^1-i \Delta^2),
\ee
which   gives
\be
E_g(1-x, \Delta^2) = -{\alpha\over  \pi} m^2 \int_0^1 d \beta {x
(1-x)^2\over (\beta (1-\beta) (\Delta^\perp)^2 (1-x)^2+ m^2 x^2)}.
\ee
If we denote the momentum fraction of the quark as $x$ instead of $1-x$, we
get,  
\be
E_g(x, \Delta^2) = -{\alpha\over  \pi} m^2 \int_0^1 d \beta {x^2
(1-x)\over (\beta (1-\beta) (\Delta^\perp)^2 x^2+ m^2 (1-x)^2)}.
\ee
In the forward limit, this gives,
\be
E_g(x,0)=-{\alpha\over \pi} {x^2\over 1-x}.
\ee
The second moment of $E_{q,g}(x,0), \int dx x E_{q,g}(x,0)$ gives in units
of ${1\over 2m}$ by how much the transverse center of momentum of the parton
$q,g$ is shifted away from the origin in the transversely polarized state.
When summed over all partons, the transverse center of momentum would still
be at the origin. Indeed it is easy to check for a dressed electron    
\be
\int_0^1 dx x E_q(x,0)+\int_0^1 dx (1-x) E_g (x,0)=0,
\ee
which is due to the fact that the anomalous gravitomagnetic moment of the
electron has to vanish. Note that in the second term, $(1-x)$ is the
momentum fraction of the gauge boson. In fact the second moment of $E_{q,g}$
appear in the angular momentum sum rule \cite{ji} and they are related to
the orbital angular momentum of the nucleon. In the approach we are
following, this connection can be seen from the fact that in the light-front
gauge, the matrix element of the quark orbital angular momentum operator
cancels the contribution of the
gluon helicity and orbital angular momentum for a dressed quark 
in the helicity sum rule \cite{oam}.

The impact parameter dependent gauge boson pdf can be  defined in the same
way as in Eq. (\ref{fteq}):
\be
{\mathcal E}_g(x, b^\perp)&=&\int {d^2 \Delta^\perp\over (2 \pi)^2}  
e^{-i b^\perp. \Delta^\perp} E_g(x,-(\Delta^\perp)^2)\nonumber\\&&
= \int {\Delta d \Delta\over (2 \pi)^2} d \theta E_g(x, -\Delta^2) 
e^{-i b \Delta cos \theta}
\nonumber\\&&=  {1\over 2 \pi}\int_0^\infty \Delta d \Delta 
E_g (x, -\Delta^2) J_0
(b \Delta).
\label{fteg}
\ee
\begin{figure}  
%\hspace{-0.3cm}
\parbox{8cm}{\epsfig{figure=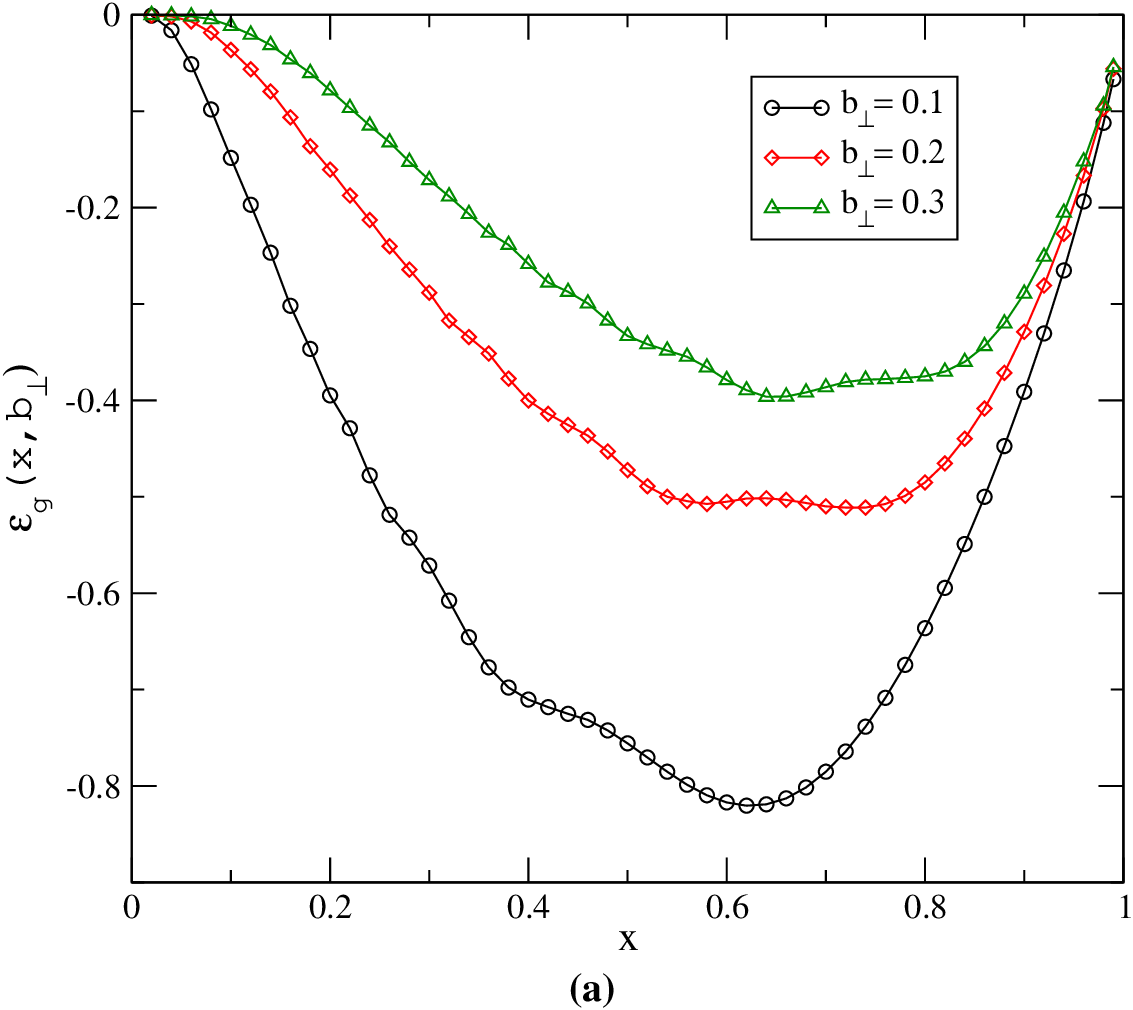,width=7.7 cm,height=7 cm}}\
\
%\hspace{0.3cm}
\parbox{8cm}{\epsfig{figure=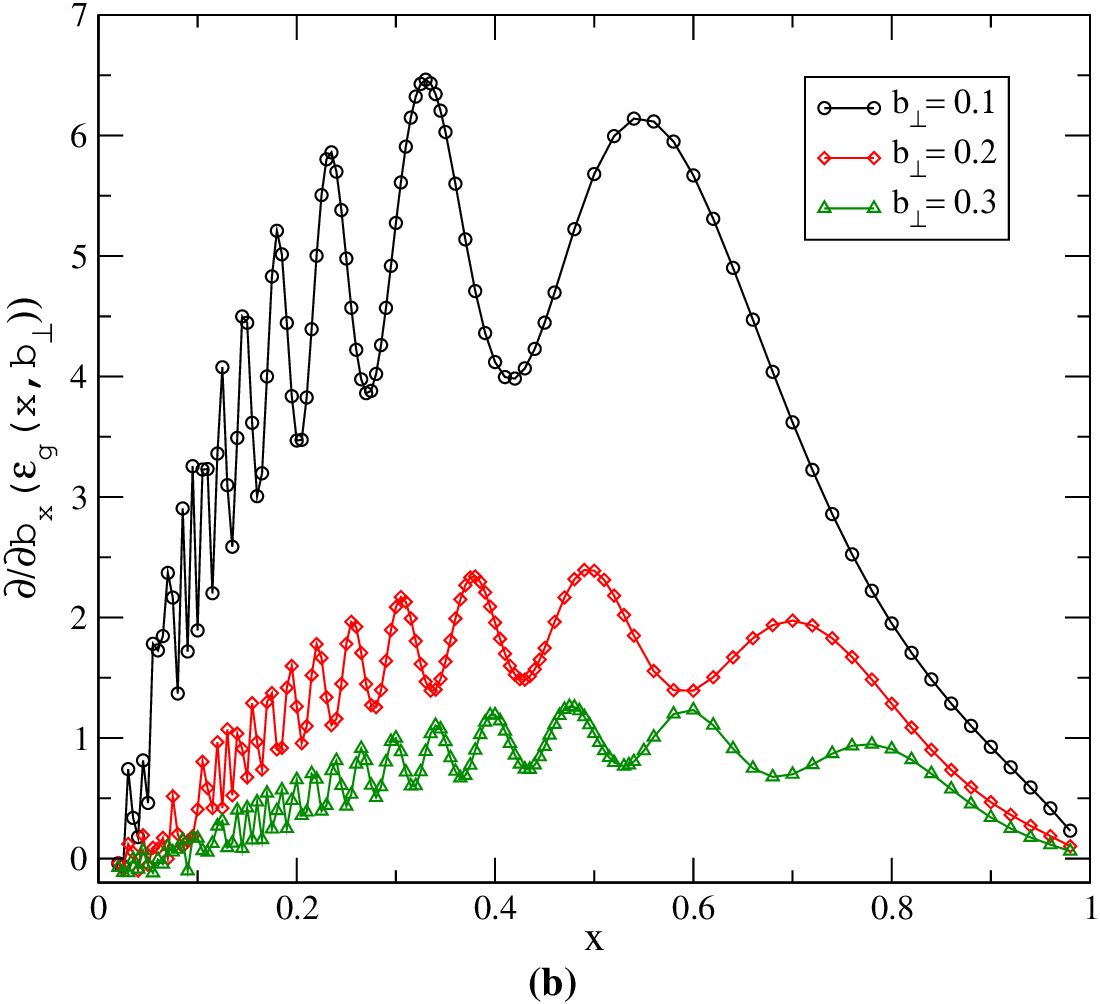,width=7.5 cm,height=7 cm}}\
\
\vspace{0.2cm} 
\begin{center} 
\parbox{14.0cm}
{{\footnotesize
Fig. 3: (Color online) (a) ${\mathcal E}_g (x, b^\perp)$ vs $x$ for three different values
of $b_\perp$; (b) Derivative of ${\mathcal E}_g (x, b^\perp)$ with respect to
$b_x$ as a function of $x$. }}
\end{center}
\end{figure}

Fig. 2 (a) shows ${\mathcal E}_g(x, b^\perp)$ vs $b^\perp$ for three
different values of $x$. ${\mathcal E}_g(x, b^\perp)$ is negative for
positive $b^\perp$ and has a negative maximum at $b^\perp=0$. As before,
 we took ${\alpha\over 2 \pi}=1$ and $m=0.5$.
 Fig. 2 (b) shows the derivative of ${\mathcal E}_g(x, b^\perp)$
with respect to $b^x$ as a function of $b^\perp$. 
 Fig. 3 (a) shows ${\mathcal E}_g(x, b^\perp)$
as a function of $x$ for three different values of $b_\perp$.
Unlike the fermion case, ${\mathcal E}_g(x, b^\perp)$ for a fixed $b^\perp$
is a complicated function of $x$ and does not increase  monotonically as $x$ 
increases. Depending on $b^\perp$, the maximum of ${\mathcal E}_g(x, b^\perp)$
appears for different $x$ ($0\le x \le 1$).
 Again in contrary to the
 fermionic case, ${\mathcal E}_g(x, b^\perp)$ becomes smoother when $x$ is
 large which is expected since  for large $x$ ($x = 0.7, 0.8$ in Fig. 2 (b)), 
gauge boson carries only a small ($1-x$) fraction of total momentum. 
 Fig. 3 (b) shows the derivative of ${\mathcal E}_g(x, b^\perp)$
with respect to $b^x$ as a function of $x$, which shows the distortion of
the pdf in the transverse plane for a transversely polarized state. 
%%%%%%%%%%%%%%%%%%%%%%%%%%%%%%%%%%%%%%%%%%%%%%%%%%%%%%%%%%%%%%%%%%%%%%
\section{Summary}
%%%%%%%%%%%%%%%%%%%%%%%%%%%%%%%%%%%%%%%%%%%%%%%%%%%%%%%%%%%%%%%%%%%%%%%%%%
We investigated the impact parameter dependent parton distributions for a
relativistic composite system. An ideal framework is based on light-front
field theory, where the transverse boosts behave like  Galilean boosts and
the longitudinal boost operator produces just a scale transformation. We
take an effective composite spin $1/2$ state, namely an electron dressed with
a photon in QED. Using the overlap representation of GPDs in terms of
light-cone wave functions, we obtained the scale dependence of the impact
parameter dependent pdfs at one loop. We also showed the
distortion of the fermion and gauge boson distributions in the transverse
plane when the state is transversely polarized. Plots show the qualitative behaviors 
of the helicity flip contributions  
${\mathcal E}_q(x, b^\perp)$ and  ${\mathcal E}_g(x, b^\perp)$ respectively 
for the electron and the gauge boson and their 
derivatives with respect to $b_x$ which give the distortion
 of the distribution 
in the transverse plane. For both fermion and gauge boson, for a fixed $x$,
the distortions are larger in magnitude for smaller $b_\perp$.

%%%%%%%%%%%%%%%%%%%%%%%%%%%%%%%%%%%%%%%%%%%%%%%%%%%%%%%%%%%%%%%%%%%%%%%%%%%
\section{acknowledgment}
%%%%%%%%%%%%%%%%%%%%%%%%%%%%%%%%%%%%%%%%%%%%%%%%%%%%%%%%%%%%%%%%%%%%%%%%%%%%

We thank M. Burkardt for valuable discussions and S.J. Brodsky for
interesting comments. We also thank the organisers of WHEPP-8 at IIT Bombay 
where part of this work was carried out. The work of AM has been supported 
in part by the 'Bundesministerium f\"ur Bildung und Forschung', Berlin/Bonn 
and the work of DC was partially suported by the Department of Energy under 
Grant No. DE-FG02-97ER-41029. 

%%%%%%%%%%%%%%%%%%%%%%%%%%%%%%%%%%%%%%%%%%%%%%%%%%%%%%%%%%%%%%%%%%%%%%%%
 
\end{document}